\documentstyle[preprint,aps]{revtex}
\begin{document}
\draft
\preprint{\vbox{Submitted to Physical Review C
                \hfill IU/NTC 94-17\\
                \null\hfill FSU/SCRI 94-124}}
\title{Relativistic nuclear structure effects in \\
quasielastic neutrino scattering}
\author{Hungchong Kim $^1$\footnote{Email : hung@iucf.indiana.edu},
J. Piekarewicz $^2$ \footnote{Email : jorgep@ds16.scri.fsu.edu},
C. J. Horowitz $^1$ \footnote{Email : charlie@iucf.indiana.edu}}
\address{$^1$ Nuclear Theory Center and Dept. of Physics,
Indiana University, Bloomington, Indiana 47408 \\
$^2$ Supercomputer Computations Research Institute, Florida State
University,\\
Tallahassee, Florida 32306}
\maketitle
\begin{abstract}
Charged-current cross sections are calculated for quasielastic
neutrino and antineutrino scattering using a relativistic
meson-nucleon model. We examine how nuclear-structure effects,
such as relativistic random-phase-approximation (RPA) corrections
and momentum-dependent nucleon self-energies, influence the extraction
of the axial form factor of the nucleon. RPA corrections are important
only at low-momentum transfers. In contrast, the momentum dependence of
the relativistic self-energies changes appreciably the value of the
axial-mass parameter, $M_A$, extracted from dipole fits to the axial
form factor. Using Brookhaven's
experimental neutrino spectrum we estimate the sensitivity of M$_A$ to
various relativistic nuclear-structure effects.
\end{abstract}
\eject
\section{INTRODUCTION}
\label{sec:intro}

The strange-quark content of the nucleon has received considerable
attention as a result of the measurement of the spin-dependent
structure function of the proton by the European Muon Collaboration
(EMC)~\cite{ashman88}. Some analyses of the experiment suggest that
a large portion of the spin of the proton is carried by strange
quarks. One can attempt to resolve this ``spin problem'' by studying
the strange-quark contribution to the vector (both electric and magnetic)
and axial-vector form factors of the nucleon. Most likely, it will take
a large number of measurements to determine all of these
form factors separately. Moreover, there are important complications
from radiative
corrections~\cite{mus}, which hinder the extraction of strange-quark
matrix elements from parity violating electron scattering. Therefore, one
anticipates a program of several
electron experiments~\cite{don92a} which, combined with
neutrino scattering data~\cite{ahr}, will offer the most accurate
strange-quark information.

Neutral-current neutrino scattering is sensitive to the
strange-quark matrix elements of the nucleon --- especially
to the isoscalar component~\cite{don92b}. Complications arise,
however, from the fact that most neutrino experiments measure a combination
of elastic scattering from free protons plus quasielastic scattering
from nucleons bound in nuclei. In the present work we examine how
nuclear-structure corrections affect the extraction of strange-quark
information.

The isoscalar part of the axial-vector form factor of the nucleon
is characterized by the parameter $g_A^s$ --- the value of the
isoscalar strange form factor at zero four-momentum transfer.
Unfortunately, the extraction of $g_A^s$ from the Brookhaven
National Laboratory (BNL) experiment is complicated by low
statistics~\cite{ahr}. Moreover, there is a strong correlation
between the extracted value of $g_A^s$ and the axial mass $M_A$
obtained from dipole
fits to the axial-vector form factor~\cite{gar,neu}. Indeed, the
world-average value of $M_A$ ($1.032 \pm 0.036 $ GeV) that the BNL
group has used is significantly different from the one
[$1.09 \pm 0.03 $ (stat)$ \pm 0.02 $ (syst)~GeV] extracted later from
a charged-current experiment~\cite{ahr1}. Thus, this large difference
in $M_A$ is sufficient to change the value of $g_A^s$ extracted from
the BNL experiment and the conclusion of a nonzero strange-quark
content from our previous neutral-current study~\cite{neu}. Therefore
determining the precise value of $M_A$ is important for strangeness studies.

At relatively high momentum transfer, the response of the nuclear
target seems to be adequately described in a relativistic Fermi gas
(RFG) model. Indeed, at a qualitative level, a RFG calculation of
quasielastic $(e,e')$ longitudinal and transverse responses agrees
well with finite-nucleus results.
However, is the RFG model accurate enough to determine the precise
value of $M_A$ ?
Currently, the BNL experiment claims an $M_A$ value with an error of
less than 5 percent.   Is there any uncertainty comparable to this small error
induced from nuclear-structure effects?  It is the aim of this work to go
beyond the RFG response and examine the sensitivity of quasielastic
neutrino scattering --- particularly $M_A$ --- to a variety of
nuclear-structure corrections.

One such correction arises from long-range RPA correlations.
In a recent paper, Singh and Oset used a nonrelativistic
RPA formalism to study the nuclear response in quasielastic neutrino
scattering~\cite{chr}. They found RPA corrections to be large only
at low-momentum transfers. Since most neutrino
experiments~\cite{expt} are carried
out at medium to high momentum transfers, they concluded that the value
of $M_A$ extracted from these experiments is reliable.

In a relativistic description of the nuclear target, additional
nuclear-structure corrections must be considered. According to
quantum hadrodynamics (QHD)~\cite{brian}, the saturation of nuclear matter
arises from a cancellation of strong scalar
($\sigma$) and vector ($\omega$) mean fields. The strong scalar field
decreases the nucleon mass while the vector field shifts the four-momentum
of the nucleon in the medium.  These relativistic effects
were not addressed in the nonrelativistic calculation and may provide
interesting corrections to the RFG response.

In a charged-current reaction, the mean-field ground state can be
characterized in terms of an effective nucleon mass that is reduced,
relative to its free-space value, by the presence of the strong
scalar field. In turn, the RPA response can be modeled from a
$(\pi + \rho + g')$ residual isovector interaction. In a mean-field
approximation, the effective nucleon mass is obtained  from solving
self-consistently the equations of motion at a given baryon
density~\cite{brian}. In the case of polarized electron scattering,
the parity-violating asymmetry was found to be sensitive to the
in-medium value of the nucleon mass~\cite{chu1}. This is an interesting
result that should be incorporated in neutrino-scattering studies.


This paper is organized as follows.   Section \ref{sec:for1} presents the
formalism for the cross section of inclusive neutrino scattering
in a relativistic impulse
approximation and in RPA.  Results for the charged current cross section are
presented in section \ref{sec:res}, while section \ref{sec:sum} is a summary.

\section{FORMALISM}
\label{sec:for1}

In a charged-current process, neutrinos (and antineutrinos) interact
with nuclei via the exchange of charged weak-vector bosons ($W^{\pm}$)
with the resulting production of charged leptons (electrons or
muons) in the final state.  In an inclusive process, where only the
final leptons are detected, the most general expression for the
cross section can be given in terms of a time-ordered product
of current operators. From this general expression several approximations
can be made depending on how one treats the ground state of the nucleus
and its response to the external probe. In Sec.~\ref{subsec:exac}, we
derive a general formalism for the inclusive process and discuss
various approximations for the response in a mean-field approximation
to the ground state. In Sec.~\ref{subsec:impulse} we discuss the form
of the nuclear current adopted in the calculation while Sec\ref{subsec:rpa}
contains a detailed description of the relativistic random phase approximation.
Finally, in Sec.~\ref{subsec:mode} we discuss how the mean-field ground state
is modified by the introduction of phenomenological momentum-dependent
corrections to  the nucleon self-energies.

\subsection{GENERAL FORMALISM}
\label{subsec:exac}

The scattering process we consider is shown in Fig.~\ref{scatt}.
An incoming neutrino with momentum $k$ scatters off the nucleus
via the exchange of weak-vector bosons producing a charged lepton
with momentum $k'$ in the final state. The initial and final states
of the nucleus are denoted by $\left| \psi_i (p_i) \right\rangle$
and $\left| \psi_f (p_f) \right\rangle$, respectively. In Born
approximation the inclusive cross section becomes proportional to
the contraction of a leptonic and a hadronic tensor:
\begin{equation}
d\sigma \propto L_{\mu \nu} W^{\mu \nu} \,
\end{equation}
where the corresponding leptonic ($L_{\mu\nu}$) and hadronic
($W^{\mu \nu}$) tensors are given by
\begin{eqnarray}
L_{\mu\nu} &=& 8 \biggl[k'_\mu k_\nu - k\cdot k' g_{\mu\nu} +k'_\nu k_\mu
\mp i\epsilon ^{\alpha\mu\beta\nu} k'^\alpha k^\beta \biggr]\label{lep} \ ,\\
W^{\mu \nu} &=& \sum_{f} \,(2 \pi)^4\,\delta^{(4)}(p_i+q-p_f)\,
\left\langle \psi_i
| {\hat J}^\mu (0) | \psi_f \right\rangle \left\langle \psi_f
| {\hat J}^\nu (0) | \psi_i \right\rangle \label{pol} .
\end{eqnarray}
Here  ${\hat J}^\mu$ is the weak charge-changing current operator of
the nucleus, $q=k-k'$ is the momentum transfer to the nucleus, and
the plus (minus) sign in $L_{\mu\nu}$ corresponds to antineutrino
(neutrino) scattering. Note that our convention for the antisymmetric
tensor is $\epsilon^{0123} \equiv 1$.

Now we introduce the current-current correlation function, or
polarization tensor, as a time-ordered product of nuclear
currents~\cite{wale}
\begin{equation}
i\Pi^{\mu \nu}(q) = \int d^4x\, e^{i q \cdot x}\,\left\langle \psi_i | T
({\hat J}^\mu (x)\,
{\hat J}^\nu (0)) | \psi_i \right\rangle  \label{corr}\ ,
\end{equation}
The hadronic tensor, and therefore the cross section, can be directly
related to the polarization tensor. In particular, it is easy to show
that the cross section takes the following form
\begin{equation}
d\sigma \propto {\rm Im}\, (L_{\mu \nu} \Pi^{\mu \nu})\ .
\end{equation}
This expression is convenient for the evaluation of the inclusive
response of a many-body system like the nucleus. In particular,
various approximations can be made depending on how one calculates
the ground state of the nucleus and its linear response to the
external probe~\cite{chu2,daw90}.

For the many-body current operator ${\hat J}_{\mu}$ we assume a
simple one-body form:
\begin{equation}
{\hat J}_\mu (x)={\bar \psi}(x) \Gamma_\mu \psi(x)\label{curr}\label{one} ,
\end{equation}
where $\psi(x)$ is a nucleon-field operator and $\Gamma_\mu$ is the
weak-interaction vertex to be discussed below [see Eq.~(\ref{cur})].
Meson-exchange currents represent corrections to this one-body form
and will be ignored throughout this paper. In a mean-field approximation
to the nuclear ground state the time-ordered product of currents
can be evaluated readily using Wick's theorem, i.e.,
\begin{equation}
i\Pi^{\mu \nu} = \int{d^4p\over (2\pi)^4}\,
  {\rm Tr}[G(p+q)\,\Gamma^\mu \,G(p)\,\Gamma^\nu] \;,
\label{tun}
\end{equation}
where $G(p)$ is the nucleon propagator that will be evaluated in
various approximations.

The simplest approximation that we employ treats the nuclear
ground state as a relativistic free Fermi gas.  Here
the nuclear response consists of the excitation of
particle-hole pairs subject to the constraints imposed by
energy-momentum conservation and the Pauli principle.
The nucleon propagator differs from the well known Feynman propagator
only because of a finite-density correction arising from the filled
Fermi sea~\cite{brian},
\begin{eqnarray}
G^o(p)=(\not\!p + M) \biggl[{1\over p^2-{M}^2 +i\epsilon} + {i\pi\over
E_{\bf p}}\delta(p_0-E_{\bf p})\ {\rm \theta}(k_F-|{\bf p}|)\biggr]
\label{gfe}\ .
\end{eqnarray}
Note that we have introduced the Fermi momentum $k_F$ and
the free (on-shell) energy $E_{\bf p}=\sqrt{{\bf p}^{2}+M^2}$.
We call this approximation ``impulse with $M$'' in order to
distinguish it from the self-consistent impulse approximation
with an effective mass $M^*$ that we now address.

One can improve the free Fermi-gas description by taking into
account, at least at the mean-field level, the interaction
between the nucleons in the nucleus. In a mean-field-theory
approximation (MFT) to the Walecka model the propagation of a
nucleon through the medium is modified by the presence of
constant scalar and vector mean-fields. These potentials
induce a shift in the mass and in the energy of a particle in the
medium and give rise to a self-consistent nucleon propagator~\cite{brian}:
\begin{eqnarray}
 G^*(p)=(\not\!{p}^{*} + M^*)
 \biggl[{1\over p^{*2}-M^{*2} +i\epsilon} +
 {i\pi \over E^*_{\bf p}} \delta({p}_{0}^{*}-E^{*}_{\bf p})
 {\rm \theta}(k_F-|{\bf p}|)\biggr] \label{gmf} \;,
\end{eqnarray}
where the effective mass and energy are shifted from their
free-space value by the scalar ($S$) and timelike component
($V^{0}$) of the mean fields,
\begin{equation}
 M^*=M+S \;; \quad E^{*}_{\bf p} =\sqrt{{\bf p^2}+M^{*2}} \;; \quad
 {p}^{*\mu}=(p^0-V^0,{\bf p}) \;.
\end{equation}
These changes in the nucleon propagator induce a corresponding change
in the polarization tensor which is now written,
\begin{equation}
i\Pi_{MF}^{\mu\nu} =  \int{d^4p\over (2\pi)^4}\, {\rm Tr}[G^*(p+q)\,\Gamma^\mu
\,G^*(p)\,\Gamma^\nu]\label{mpol}\ .
\end{equation}
Note, in computing the response we
integrate over the four-momentum of the nucleons, and the contribution from
the constant vector potential can be eliminated by a simple change of
variables. [This will not happen once momentum-dependent corrections are
incorporated into the mean fields (see Sec.~\ref{subsec:mode}).] Formally
the mean-field response is identical to that of a relativistic Fermi
gas of nucleons with an effective mass $M^*$. We refer to this calculation
as ``impulse with $M^*$''.

\subsection{IMPULSE APPROXIMATION}
\label{subsec:impulse}

We start this section by writing the inclusive cross section
--- per neutron --- for the charge-changing process as
\begin{equation}
{d^2\sigma\over {d\Omega _{k'} d E_\mu}}=-{G_F^2\, {\rm cos^2\theta _c}\,
|{\bf k}'|\over
{32 \pi^3 \rho E_\nu}}\,{{\rm Im}\,(L_{\mu\nu} \Pi^{\mu\nu})}\ .
\label{e1}
\end{equation}
Here $\rho = k_F^3/3\pi^2$ is the neutron (or proton) density of the
system, ${\rm \theta}_c$ the Cabbibo angle (${\rm cos}^2\theta_c=0.95$),
$G_F$\, is the Fermi constant, ${\bf k}'$\, the three-momentum of the
outgoing lepton, and $E_\nu$ the energy of the incoming neutrino
(or antineutrino).

In the impulse approximation the interaction between the incoming
neutrino and a target nucleon is assumed to be the same as in free
space. Hence, we employ a charge-changing current operator
with single-nucleon form factors parameterized from on-shell data.
That is (suppressing isospin labels),
\begin{equation}
\Gamma^\mu(q)= F_1(Q^2)\gamma^\mu +
              iF_2(Q^2) \sigma^{\mu\nu} {q_\nu \over 2M} -
               G_A(Q^2)\gamma^\mu \gamma^5 +
               F_p(Q^2)q^\mu\gamma^5\  \;, \quad
               (Q^2 \equiv {\bf q}^{2}-q_{0}^{2})
\label{cur}
\end{equation}
The form factors $F_1, F_2, G_A$ and $F_p$ are given in Appendix~\*{1}.  The
pseudoscalar form factor $F_p$ is constructed from PCAC and its contribution
is suppressed by the small lepton mass.

Since $\Gamma^{\mu}$ has been expressed in terms of vector, tensor,
axial-vector, and pseudoscalar vertices, the inclusive cross section
requires the evaluation of a large set of nuclear-response functions.
These are conveniently separated in the following way (note that the
subscripts indicate the vertices involved):
\begin{eqnarray}
\Pi_{vv} ^{\mu\nu}&=&-i \int{d^4p\over (2\pi)^4}\,
{\rm Tr}[G(p+q)\,\gamma^\mu \,G(p)\,
\gamma^\nu]\label{pvv}\ ,\\
\Pi_{tt} ^{\mu\nu}&=&-i \int{d^4p\over (2\pi)^4}\,
{\rm Tr}[G(p+q)\,{i\sigma^{\mu\alpha}q_\alpha\over 2M} \,G(p)\,
{-i\sigma^{\nu\beta}q_\beta\over 2M}]\label{ptt}\ ,\\
\Pi_{vt} ^{\mu\nu} &=& -i \int{d^4p\over (2\pi)^4}\,
{\rm Tr}[G(p+q)\,\gamma^\mu \,G(p)\,
{-i\sigma^{\nu\beta}q_\beta\over 2M}]\label{pvt}\ ,\\
\Pi_{va} ^{\mu\nu} &=& -i \int{d^4p\over (2\pi)^4}\,
{\rm Tr}[G(p+q)\,\gamma^\mu \,G(p)\,
\gamma^\nu\gamma^5] = -
i \epsilon^{\mu \nu \alpha 0}\, q_\alpha \Pi_{va}\label{pva}
\ ,\\
\Pi_{aa} ^{\mu\nu} &=& -i \int{d^4p\over (2\pi)^4}\,
{\rm Tr}[G(p+q)\,\gamma^\mu\gamma^5 \,G(p)\,
\gamma^\nu\gamma^5] = \Pi_{vv} ^{\mu\nu}+g^{\mu\nu}\Pi_{A}\ ,\\
\Pi_{ap} ^{\mu\nu} &=& -i \int{d^4p\over (2\pi)^4}\,
{\rm Tr}[G(p+q)\,\gamma^\mu\gamma^5 \,G(p)\,
(-q^\nu)\gamma^5] = q^\mu q^\nu \Pi_{ap}\ ,\\
\Pi_{pp} ^{\mu\nu} &=& -i \int{d^4p\over (2\pi)^4}\,
{\rm Tr}[G(p+q)\,q^\mu\gamma^5 \,G(p)\,
(-q^\nu)\gamma^5]= q^\mu q^\nu \Pi_{pp}\ .
\end{eqnarray}
The various components of the polarization tensor
can be computed in the Fermi-gas limit or in the mean-field
approximation by using either $G^o(p)$ [Eq.~(\ref{gfe})]
or $G^*(p)$ [Eq.~(\ref{gmf})] respectively. The imaginary parts
of all these polarizations have been calculated analytically following the
Ref.~\cite{ho84} and are given in Appendix~\*{2}.

The polarizations $\Pi_{vv} ^{\mu\nu}$,\, $\Pi_{tt} ^{\mu\nu}$,\,
and $\Pi_{vt} ^{\mu\nu}$\ are only sensitive to the Lorentz-vector
part of the weak current and, thus, satisfy current conservation,
$q_{\mu}\Pi^{\mu\nu}=\Pi^{\mu\nu}q_{\nu}=0$. The conservation
of the vector current plus Lorentz covariance imply that only two
components of each of these polarizations are independent. These
have been chosen to be the longitudinal and transverse components
which are defined, for example in the the case of $\Pi_{vv} ^{\mu\nu}$,
as
\begin{eqnarray}
   \Pi^L_{vv} &\equiv&   \Pi^{00}_{vv}-\Pi^{11}_{vv} = -
                     {q^2 \over {\bf q}^2} \Pi^{00}_{vv}    \;, \\
   \Pi^T_{vv} &\equiv& 2 \Pi^{22}_{vv} = 2 \Pi^{33}_{vv} \;.
\end{eqnarray}
Here we have assumed a coordinate system with the
$\hat{x}=\hat{1}$ axis along the direction of the
three-momentum transfer ${\bf q}$. Using Lorentz
covariance one can isolate the additional responses
that arise from the other components of the current
and obtain the following invariant
amplitude
\begin{equation}
L_{\mu\nu}\Pi^{\mu\nu} = L_L R_L + L_T R_T \pm L_{va} R_{VA}
+L_A R_A + L_p R_P\label{inva}\ .
\end{equation}
The nuclear-structure information is fully contained in the various
response functions which have been defined in terms of the above
polarization tensors:
\begin{eqnarray}
R_L &=& (F_1^2 + G_A^2 )\ \Pi^L_{vv} +
2 F_1 F_2\ \Pi^L_{vt}+F_2^2\ \Pi^L_{tt}\ ,\\
R_T &=& {1 \over 2}[(F_1^2 + G_A^2 )\ \Pi^T_{vv} +
2 F_1 F_2\ \Pi^T_{vt}+F_2^2\ \Pi^T_{tt}]\ ,\\
R_P &=& 2G_A\ F_p\ \Pi_{ap}+F_p^2\ \Pi_{pp}\ ,\\
R_A &=& G_A^2 \Pi_A \ ,\\
R_{VA} &=& (F_1+F_2 {M^* \over M})\ G_A\ |{\bf q}|\ \Pi_{va}\ .
\end{eqnarray}
These response functions are multiplied by appropriate
kinematical factors that could, in principle, be used to separate
the individual responses
\begin{eqnarray}
L_L&\equiv&-{q^2 \over {\bf q}^2} L_{00} -{4 m_\mu^2\ q_0 \over {\bf q}^2}
(4 E_\nu -q_0-{q_0\ m_\mu^2 \over q^2})\ ,\\
L_A&\equiv&8(q^2-m_\mu^2)\ ,\\
L_T&\equiv&-{q^2 \over {\bf q}^2} L_{00}-{4 m_\mu^2 \over {\bf q}^2}
(4 E_\nu\ q^0-q^2+ m_\mu^2)-L_A\ ,\\
L_p&\equiv&4 m_\mu^2\ ( m_\mu^2-q^2)=-{1\over 2}m_\mu^2\ L_A\ ,\\
L_{va}&\equiv&-16{q^2\ (E_\nu+E_\mu)+q_0\ m_\mu^2 \over |{\bf q}|} \ .
\end{eqnarray}
Finally, we note that the plus (minus) sign in Eq.~(\ref{inva}) should be
used for neutrino (antineutrino) scattering.

\subsection{RELATIVISTIC RANDOM PHASE APPROXIMATION}
\label{subsec:rpa}
In the present section we improve the simple particle-hole description of the
response by incorporating many-body RPA correlations. Many-body correlations
can be included by considering the residual interaction between the
particle and the hole. For the present charge-changing reaction only
isovector correlations are important. Yet, there might still be
important effects associated with the isoscalar mean fields (e.g., $M^*$).
Indeed, in a recent calculation we have shown that the reduction of
the effective nucleon mass in the medium results in a quenching of the
effective $NN\pi$ coupling which, in turn, is responsible for suppressing
the predicted enhancement of the spin-longitudinal to spin-transverse ratio,
in accordance with experiment~\cite{chu3}. For the residual isovector
interaction we employ a simple relativistic generalization of the
conventional
$\pi + \rho + g'$ interaction~\cite{chr,eng}. The phenomenological
Landau-Migdal
parameter $g'$ has been included to simulate the effect of
repulsive short-range
correlations. The RPA correction, $\Delta \Pi^{\mu\nu}_{RPA}$, to
the polarization
tensor, $\Pi^{\mu\nu}$ in Eq.~(\ref{e1}), is shown diagrammatically in
Fig.~\ref{rpafig}~(a). The RPA corrections are calculated from
the dressed propagator, ${\bf D_{RPA}}$.  This includes an
infinite sum of the lowest-order (uncorrelated) polarization
as illustrated in Fig.~\ref{rpafig}~(b).  Note, for the mean-field
ground state, the nucleon propagators in the lowest-order
polarization already
include the isoscalar dressing due to the mean fields [see Eq.~(\ref{gmf})].


The Lagrangian density describing the isovector component of the NN
interaction is given by
\begin{equation}
{\cal L}= g_\rho {\bar \psi} \gamma^\mu
{\mbox{\boldmath $\tau$} \over 2}
\cdot \psi
\mbox{\boldmath $\rho$}_\mu +
f_\rho {\bar \psi} \sigma^{\mu \nu}
{\mbox{\boldmath $\tau$} \over 2} \cdot \psi
{\partial_\mu
\over 2M} \mbox{\boldmath $\rho$}_\nu -
{f_\pi \over m_\pi} {\bar \psi} \gamma_5 \gamma^\mu
\mbox{\boldmath $\tau$} \psi \cdot \partial_{\mu}
\mbox{\boldmath $\pi$}\label{inte}\ .
\end{equation}
The $\rho$ meson has a vector ($g_{\rho}$) as well as a tensor
($f_{\rho}$) coupling to the nucleon. The parameters of the
model are obtained directly from the Bonn potential fit to NN
properties and are given by $g_{\rho}^{2}/4\pi=1.64$ and
$f_{\rho}/g_{\rho}=6.1$~\cite{bonn} (note that our value
for $g_{\rho}^2/4\pi$ is four times larger than the one
quoted by the Bonn group simply because of our selection
of \mbox{\boldmath $\tau$}/2, rather than \mbox{\boldmath $\tau$},
as the isospin vertex). With this form for the interaction
Lagrangian the $NN\rho$ vertex becomes (combined with isospin
matrices)
\begin{equation}
  \Gamma_{NN\rho} = -{i \over \sqrt{2}}
   \biggl(g_\rho \gamma_\mu +
   {if_\rho\sigma_{\mu\nu} q^\nu\over 2M}\biggr)\ ,
\end{equation}
while the $\rho-$meson propagator is given by
\begin{equation}
  R_{\mu\nu}(q)= {-g_{\mu\nu} + q_{\mu}q_{\nu}/m_{\rho}^{2}
  \over q^2-m_\rho^2+i\epsilon}\ .
\label{rhoprop}
\end{equation}
Note that the ``gauge'' piece $q_{\mu}q_{\nu}/m_{\rho}^{2}$
will not contribute to the RPA response because in the
mean-field approximation the vector-isovector current is
conserved.
For the $NN\pi$ vertex we have adopted a pseudovector form with
$f_{\pi}^2/4\pi=0.075$. The pion, rho, and free nucleon masses
have been fixed at their experimental values.

The most uncertain parameter in our calculation is the
phenomenological Landau-Migdal parameter $g'$. Traditionally,
$g'$ is introduced to regularize the large spin-spin component
of the isovector interaction. In a relativistic formalism we
can incorporate short-range correlation effects by modifying
the pion ``propagator'' in the following way:
\begin{equation}
  V_{\mu\nu}=  {q_\mu q_\nu \over {q^2-m_\pi^2+i\epsilon}} \rightarrow
  V_{\mu\nu}=  {q_\mu q_\nu \over {q^2-m_\pi^2+i\epsilon}} - g'g_{\mu\nu}\ .
\label{pio}
\end{equation}
Without $g'$ the contribution of the pion to the RPA response would
be suppressed by current conservation and the small leptonic mass.
Therefore, it is through the Landau-Migdal parameter $g'$ that the
pion mainly contributes to the RPA response (note that the term
proportional to $g'$ has, both, longitudinal as well as transverse
components). With this choice for the pion propagator the ``elementary''
$NN\pi$ vertex becomes
\begin{equation}
   \Gamma_{NN\pi}=\sqrt{2} {f_{\pi} \over m_{\pi}} \gamma^5 \gamma^\mu\ .
\end{equation}

With the above Feynman rules in hand we can now construct the
medium-modified rho- and pion-mediated interactions. Note
that the inclusion of $g'$, which contains transverse as well
as longitudinal components, is responsible for $\rho-\pi$ mixing.
In order to properly account for this mixing, Dyson's equation for
the propagator must be expanded from a $4\times 4$ to an $8\times 8$
matrix equation:
\begin{equation}
 {\bf D}_{RPA} = {\bf D}_{0} + {\bf D}_{0} {\bf \Pi}_{0} {\bf D}_{RPA}\ ,
\end{equation}
where we have defined the free (diagonal) propagator matrix
\begin{equation}
{\bf D}_{0} =  \left( \begin{array}{cc}
              {\bf R} & {\bf 0 }\\
              {\bf 0} & {\bf V}
               \end{array} \right) \;,
\end{equation}
in terms of the $\rho$ [Eq.~(\ref{rhoprop})] and $g'-$modified
pion propagator [Eq.~(\ref{pio})]. We have also introduced the
mixed $\rho-\pi$ polarization matrix
\begin{equation}
{\bf \Pi}_{0} = \left( \begin{array}{cc}
             {\bf \Pi}_{\rho\rho} & {\bf \Pi}_{\rho\pi} \\
             {\bf \Pi}_{\pi\rho} & {\bf \Pi}_{\pi\pi}
             \end{array} \right) \ ,
\end{equation}
with individual components given by
\begin{eqnarray}
\Pi_{\rho\rho}^{\mu\nu} &=& -{i\over 2} \int{d^4p\over (2\pi)^4}\,
{\rm Tr}\Biggl[\biggl(g_\rho\gamma^\mu+{if_\rho
\sigma^{\mu\alpha}q_\alpha\over2M}\biggr)G(p)\biggl(g_\rho\gamma^\nu-{if_\rho
\sigma^{\nu\beta}q_\beta\over2M}\biggr)G(p+q)\Biggr]\ ,\\
\Pi_{\rho\pi}^{\mu\nu} &=& \Pi_{\pi\rho}^{\mu\nu}= -{ig_\rho f_\pi\over m_\pi}
\int{d^4p\over (2\pi)^4}\,{\rm Tr}\Biggl[\biggl(\gamma^\mu+{if_\rho
\sigma^{\mu\alpha}q_\alpha\over2Mg_\rho}\biggr)
G(p)\gamma^5\gamma^\nu G(p+q)\Biggl]\ ,\\
\Pi_{\pi\pi}^{\mu\nu} &=& -{i2f_\pi^2\over m_\pi^2} \int{d^4p\over (2\pi)^4}\
{\rm Tr}\biggl[\gamma^5\gamma^\mu G(p)\gamma^5\gamma^\nu G(p+q)\biggl]\ .\\
\end{eqnarray}

The RPA correction to the polarization now takes the following form
[see Fig.~\ref{rpafig}~(a)]:
\begin{eqnarray}
{\bf \Delta\Pi}_{RPA}=({\bf \Pi}_\rho, {\bf \Pi}_\pi)\,
{\bf D}_{RPA} \left(
\begin{array}{c} {\bf \Pi}_\rho\\ {\bf \Pi}_\pi \end{array} \right)\ ,
\end{eqnarray}
where ${\bf \Pi}_\rho$ and ${\bf \Pi}_\pi$ characterize the in-medium
mixing --- due to particle-hole excitations ---- of a charged weak-vector
boson with a $\rho$ or $\pi$ meson, respectively, and are given by
\begin{eqnarray}
\Pi_\rho^{\mu\nu} &=& - i{g_\rho \over \sqrt{2}} \int{d^4p\over (2\pi)^4}\,
{\rm Tr}\Biggl[\Gamma^\mu G(p)(\gamma^\nu-{if_\rho
\sigma^{\nu\alpha}q_\alpha\over
2Mg_\rho})G(p+q)\Biggr]\ ,\\
\Pi_\pi^{\mu\nu} &=& - \sqrt{2} i {f_\pi\over m_\pi} \int{d^4p\over
(2\pi)^4}\,{\rm
Tr}\Biggl[\Gamma^\mu G(p)\gamma^5\gamma^\nu G(p+q)\Biggr]\ .
\end{eqnarray}
An RPA calculation of the inclusive response uses the same
expression for the cross section as in Eq.~(\ref{e1}) with
the replacement:
\begin{equation}
\Pi ^{\mu\nu} \rightarrow \Pi ^{\mu\nu}_{RPA}=
\Pi ^{\mu\nu}+\Delta\Pi_{RPA}^{\mu\nu} \;.
\end{equation}

In an impulse (or uncorrelated) description of the response
the cross section is only sensitive to the imaginary part of the lowest-order
polarizations. Since the nuclear response is being probed in
the spacelike ($q^{2}<0$) region, $N\bar{N}$ pairs can not be
excited in these lowest-order descriptions. They can, however,
be virtually excited and, thus, will become an essential
ingredient of the RPA response. Traditionally, $N\bar{N}$
excitations have been divided into two contributions, one
being vacuum polarization and the other consisting of the Pauli
blocking of $N\bar{N}$ excitations due to the filled Fermi
sea~\cite{chu2}. The latter contribution is finite and has been
shown to be essential for the conservation of the electromagnetic
current. This contribution has been included in our calculations.
The former, however, is divergent and must be renormalized. Since
we are using a nonrenormalizable theory with derivative couplings,
the renormalization of these divergent contributions becomes
ambiguous at best. Thus, in order to avoid including ad-hoc
parameters (e.g., cut-offs) we have decided to simply
ignore the effect from vacuum polarization. Note that the
(finite) real parts of the various polarizations have
been calculated analytically and most of them have been
published already~\cite{lim}.  Here we calculate
them numerically so we can extend the formalism to include
momentum-dependent self-energies.

\subsection{Momentum dependent vector and scalar self energy}
\label{subsec:mode}

In a mean-field approximation to the Walecka model the
vector ($V$) and scalar ($S$) self-energies are replaced
by their classical expectation values. In this approximation
the nucleon self-energy is real and energy independent.
However, as the momentum of the nucleon becomes large
there is an important coupling of the nucleon to nuclear
excitations. Indeed, at intermediate energies it is known
that the reactive content of the reaction is dominated by
quasifree nucleon knockout. Thus, at large-enough momentum
the nucleon self-energy will become complex and energy
dependent. In order to calculate the nuclear response for
a broad range of momentum transfers, we incorporate
momentum-dependent self-energies into the nucleon propagator.
Since a microscopic calculation of the energy dependence of
the nucleon self-energy in the Walecka model awaits, we have
used the phenomenological optical potentials of Ref.~\cite{bc1}.
A detailed discussion of this momentum-dependent correction
can be found in Ref.~\cite{ele}.

In a calculation with momentum-dependent self-energies,
the effective mass and energy of nucleons in the medium
are no longer equal. In particular, the nucleon propagator
must now be described in terms of Dirac spinors with masses
and energies given by
\begin{equation}
  M_{\bf p}^{*}=M+S({\bf p}) \;; \quad
  E_{\bf p}=E_{\bf p}^* + V({\bf p}) =
  \sqrt{{\bf p^2}+M_{\bf p}^{*2}} + V({\bf p}) \;.
\end{equation}
The basic formalism, however, remains unchanged except
for the inclusion of a more realistic nucleon propagator
given by
\begin{equation}
G(p) = (\not\!p^* + M_{\bf p}^{*}) \biggl[{1\over
p^{*2}-{M_{\bf p}^{*2}} +i\epsilon} +
{i\pi\over
E_{\bf p}^*}\delta(p_{0}^{*}-E_{\bf p}^{*}){\rm \theta}(k_F-|{\bf p}|)\biggr]\,
\end{equation}
where $p^{* \mu} = (p^0-V({\bf p}), {\bf p})$\ .
Note that the vector potential can no longer be eliminated from
the integrals defining the polarization by a simple change of
variables.

\section{RESULTS}
\label{sec:res}

In this section we present results for the inclusive cross section using
a variety of approximations. We consider impulse-approximation calculations
using, both, a relativistic Fermi gas of nucleons of mass $M$ and
a self-consistent ground state with an effective nucleon mass of $M^{*}$.
We also present two calculations including
RPA correlations either with mass $M$\ or $M^*$.   The mass $M$\ RPA
calculations can be directly compared to similar nonrelativistic calculations.
Finally, results will be shown
using momentum-dependent self-energies obtained from the
phenomenological fit to the
nucleon optical potential of Ref.~\cite{bc1}. We refer to this
last calculation as
``impulse with $M_{\bf p}^{*}$''. The impulse with $M$ calculation
is commonly used to extract from experiment the mass parameter $M_A$
present in the dipole fit to the
axial form factor $G_{A}$. Our main goal is to estimate the
sensitivity of this
parameter to various relativistic nuclear-structure effects.

The effective nucleon mass $M^{*}$ (in mean field theory) is
obtained from a solution to the self-consistency equation~\cite{brian},
\begin{equation}
M^*= M - {g_s^2 \over m_{s}^{2}} {4 \over (2\pi)^3}
 \int^{k_F}_{0} d^3 k {M^* \over
\sqrt{{\bf k}^2+{M^*}^2}}\ .
\end{equation}
Choosing the couplings to reproduce the bulk properties of nuclear
matter at saturation [$g_s^2(M^{2}/m_{s}^{2})=267.1$] leads to a
value of the effective nucleon mass of $M^*=638$~MeV at an assumed
average density corresponding to $k_F=225$ MeV.

In Fig.~\ref{qu1} we show the double differential cross section
$d^2\sigma/d\Omega_{k'} dE_\mu$ [see Eq.~(\ref{e1})] for a
neutrino energy of $E_\nu=1$ GeV. Fig.~\ref{qu1}~(a) is for a
momentum transfer of $|{\bf q}|=0.5$~GeV while Fig.~\ref{qu1} (b) is
for $|{\bf q}|=1.2$~GeV. At $|{\bf q}|=0.5$~GeV the peak positions
from the $M^*$ and $M_{\bf p}^{*}$ calculations are shifted by less
than 50 MeV relative to the Fermi-gas peak. This value represents
an average binding-energy shift and has been observed experimentally
in quasielastic electron scattering. Thus, a simple mean-field
calculation is expected to give a reasonable description of the
nuclear response. The situation changes considerably, however, at
$|{\bf q}|=1.2$~GeV [Fig.~\ref{qu1} (b)]. Here, the $M^{*}$
calculation predicts a shift in the peak position that is substantially
larger than the one obtained using the more realistic ($M_{\bf p}^{*}$)
self-energies. Note that the impulse with $M_{\bf p}^{*}$ calculation
still predicts a substantial shift (of the order of 50 MeV) relative
to the Fermi-gas value.
Also note the kinematical cutoff in $q_{0}$
--- above this value the scattering is prohibited kinematically. The impulse
with $M^{*}$ calculation has a considerable amount of strength shifted
into this inaccessible region.
At this momentum transfer, the mean-field calculation overpredicts
the binding energy shift seen in electron scattering\cite{ele}.
In contrast, the calculation using the impulse with $M_{\bf p}^{*}$
shows a reasonable binding-energy shift and is practically insensitive
to the kinematical cutoff.

Fig.~\ref{qu2} shows the inclusive RPA cross section at $|{\bf q}|=0.5$~GeV
for various values of the Landau-Migdal parameter $g'$. For reference, the
impulse results with $M$ and with $M^{*}$ have also been included in
Fig.~\ref{qu2}~(c). The solid line shows the RPA response of a
Fermi gas ground state while the dashed line shows the RPA
response of the mean-field ground state.
Since the pion contributes only through $g'$, the softening
and enhancement of
the response in Fig.~\ref{qu2}~(a) is exclusively
due to an attractive --- and thus unrealistic --- rho-mediated residual
interaction. As the value of $g'$ is
increased to $g'=0.3$, the peak moves to higher excitation
energy and the overall
strength of the response is reduced. This is consistent with an additional
repulsive component to the residual interaction arising from $g'$. Finally,
Fig.~\ref{qu2}~(c) shows the RPA results using the standard value of $g'=0.7$.
One now observes a slight quenching and hardening of the response due to the
large value of $g'$.

Since most of the experiments report the energy-integrated cross section
$d\sigma/dQ^2$ we have taken our results for the double differential
cross section and integrated over the allowed kinematical region
of $q_0$
\begin{equation}
{d\sigma\over dQ^2} = \int^{Q_c}_{0} {\pi \over E_\nu k'} {d^2 \sigma \over
d E_\mu d\Omega} dq_0\ ,
\end {equation}
where the energy cutoff is given by
\begin{equation}
Q_c = E_\nu+{q^2-m^2_\mu \over 4 E_\nu }+
{E_\nu m^2_\mu\over q^2 - m^2_\mu}\ ,
\end {equation}
and $m_\mu$ is the mass of the produced lepton.
In Fig.~\ref{Neu}~(a) the energy-integrated cross section, $d\sigma/dQ^2$,
at $E_\nu=1.2 $~GeV is shown for three calculations: impulse with $M$,
RPA with $M^*$ using $g'=0.7$, and impulse with $M_{\bf p}^{*}$.
Note that there is a substantial quenching due to RPA correlations
at small momentum transfers, i.e., $Q^2 \leq 0.3$ GeV$^2$. In the
intermediate range ($0.3 \leq Q^2 \leq 0.9$ GeV$^2$) the importance
of RPA correlations diminishes and no significant differences are
observed between the three models. At an even larger $Q^2$ the RPA
curve splits from the other two indicating the breakdown of the
mean-field approximation [see Fig.~\ref{qu1} (b)]. This breakdown,
however, is not associated to RPA effects --- which are no longer
effective at this momentum transfer --- but rather from
the mean-field $M^*$ effects. Indeed, an RPA calculation using
the free nucleon mass is within one percent of the impulse with $M$
calculation at high $Q^{2}$.

We report similar calculations for antineutrinos in Fig.~\ref{Neu} (b).
The cross sections are considerably reduced relative to
neutrinos because of the sign change in the vector-axial
interference term [see Eq.~(\ref{inva})]. In addition, since the cross
sections fall to (almost) zero for $Q^2 \geq 1$~GeV$^2$, most of the
quasifree strength is located below the cutoff $Q_c$. Hence, the
kinematical cutoff does not have a significant effect on the $M^*$ curve
at large $Q^{2}$. At a smaller momentum transfer, $Q^2 \leq 0.3$~GeV$^2$,
and just as in the neutrino case, RPA correlations substantially reduce the
cross section. At a larger $Q^{2}$, most of the differences observed between
the RPA and the impulse with $M$ calculations are due to the strong scalar
potential. The impulse with $M_{\bf p}^{*}$ calculation (dot-dashed curve)
smoothly interpolates these two models.

In the BNL experiment~\cite{ahr1}, the axial mass $M_A$, which controls
the $Q^{2}$ falloff of the axial form factor [see Eq.~(\ref{gax})], was
extracted from antineutrino data using
a Fermi gas (i.e., impulse with $M$) formalism. From an analysis of their
data in the range $Q^2=0.2-1$~GeV$^2$, the BNL group extracted a value
of $M_A=1.09 \pm 0.03 $ (stat)$\ \pm 0.02 $\ (syst)~GeV. In Fig.~\ref{bnlfig}
we show the cross section, $d\sigma/dQ^2$, averaged over the BNL neutrino
spectrum. In Fig.~\ref{bnlfig}~(a) we present four different calculations:
impulse and RPA with $M$ (solid lines), and impulse and RPA with $M^*$
(dashed lines). The two curves using the free nucleon mass $M$ start to
overlap at $Q^2=0.4$~GeV$^2$ and, insofar as we can regard them as our
nonrelativistic limit, they agree with the nonrelativistic results
obtained by Singh and Oset~\cite{chr}. Similarly, the two $M^*$ curves
also coincide for $Q^2 \geq 0.4$~GeV$^2$  illustrating the fact that RPA
correlations become unimportant in this $Q^{2}$ region.
In Fig.~\ref{bnlfig}~(b) we show three impulse curves in the
$Q^2=0.2-1$~GeV$^2$  range --- the region sampled in the BNL experiment.
At $Q^2 =0.2$ there is a 6 percent difference between the impulse
with $M$ calculation and the $M^*$ and $M_{\bf p}^{*}$ values.
This difference becomes larger, 20--30 percent, at $Q^2=1.0$~GeV$^2$.
It is, therefore, essential to estimate the sensitivity of $M_{A}$
to these nuclear-structure effects.

The raw experimental data
as a function of the four-momentum transfer $Q^2$\ was
fitted with a theoretical curve to determine $M_A$. Since the experiment
suffers from an uncertainty
in the overall normalization, we use the ratio of the cross section at two
different values of $Q^{2}$ to extract $M_{A}$. In Table~\ref{compa} we show
the ratio of the cross section using $Q^2=0.2$~GeV$^2$ and $Q^2=1.0$~GeV$^2$.
The value of the axial-mass parameter $M_{A}$ was varied
in the $1.09-1.30$~GeV
range and the ratio of cross sections reported for
various nuclear-structure models.
For $M_A=1.09$~GeV --- the value extracted
from the BNL data --- the impulse with
$M$ ratio equals 19.42. This represents our baseline value for the ratio. In
order to reproduce this value using the impulse with $M^{*}$
calculation the value
of the axial-mass must be changed to
$M_A=1.30$~GeV (a 20 percent increase). Similarly,
the value of $M_{A}$ must be changed to 1.25 GeV in the RPA with $M^*$ and to
1.20 GeV in the $M_{\bf p}^{*}$ calculation. There is, however, no $M_A$ value
within this range that can reproduce the
ratio using an RPA with $M$ calculation.
This is because RPA effects substantially reduces
the cross section at $Q^2=0.2$
(about 10 percent) which was not expected from nonrelativistic calculations.

Since we have established that RPA correlations become unimportant for
$Q^2 \geq 0.4$~GeV$^2$, we can eliminate the sensitivity to RPA effects
by computing the ratio of cross sections using $Q^2=0.5$~GeV$^2$ and
$Q^2=1.0$~GeV$^2$ (see Table~\ref{compa1}). In this case, using the
impulse with $M$ calculation we obtain a baseline value for the ratio
of 5.47 at $M_A=1.09$~GeV. Now the RPA with $M$ does not induce any
change in the value of $M_A$  --- in agreement with the
nonrelativistic calculation of Singh and Oset~\cite{chr}. This,
however, is not the case for the other nuclear-structure calculations.
Indeed, even in the impulse with $M_{\bf p}^{*}$ calculation --- which
generates the smallest change --- a 10 percent increase in the value of
$M_{A}$ is required. This 10 percent uncertainty in the value of $M_A$
could complicate the extraction of strange-quark matrix elements from
experimental studies of neutral weak form factors~\cite{neu} .

\section{CONCLUSIONS AND OUTLOOK}
\label{sec:sum}

We have used a relativistic formalism to study inclusive charged-current
neutrino scattering in an impulse approximation. The nuclear-structure
information is contained in a large set of nuclear-response functions that
were evaluated in nuclear matter using a variety of approximations. The
simplest approximation that we considered was a relativistic free Fermi gas.
This approximation was used by the BNL group to extract the axial mass
parameter $M_{A}$. We have used the Fermi-gas approximation --- together
with the BNL-extracted value of $M_{A}=1.09$~GeV --- to fix the $Q^{2}$
dependence of the inclusive cross section. We have examined the
sensitivity of this extracted value for $M_{A}$ to various nuclear-structure
effects --- such as the ones arising from the mean fields ($M^{*}$), RPA
correlations, and momentum-dependent self-energies. In essence, we have
computed the changes in $M_{A}$ that were required to reproduce the baseline
free Fermi gas cross section once these additional
nuclear-structure effects were taken into
consideration. This analysis is useful because a robust value of $M_{A}$ is
essential for a reliable determination of the strange-quark content of the
nucleon.

Our results indicate important corrections to the nuclear response due
to the mean fields ($M^{*}$) and to RPA correlations. Indeed, changes
as large as 20 percent in $M_{A}$ were observed whenever the whole range
of $Q^{2}$ values ($0.2~{\rm GeV}^{2} \le Q^{2} \le 1.0~{\rm GeV}^{2}$)
used in the BNL experiment were employed for the extraction. This
uncertainty, however, can be substantially reduced. For example,
it is well known that RPA correlations are effective only at small
momentum transfers (i.e., $Q^{2} \le 0.3$~GeV$^{2}$). In addition,
phenomenological fits to the nucleon optical potential indicate that,
at large nucleon momenta, the mean fields are considerably smaller in
magnitude than the ones predicted by the mean-field theory. Hence, some
of the nuclear-structure uncertainties can be removed by employing
phenomenological (momentum-dependent) mean fields and by restricting the
range of $Q^{2}$ to the region
$0.5~{\rm GeV}^{2} \le Q^{2} \le 1.0~{\rm GeV}^{2}$.
Note that, although weaker than in the mean-field theory, the
momentum-dependent optical potentials are still large and induce
nontrivial changes in the effective mass and energy of a particle in the
medium. Indeed, even in this best-case scenario a 10 percent
uncertainty in $M_{A}$
persists.  With this extra 10 percent uncertainty, the BNL
experiment~\cite{ahr} by itself no longer provides strong evidence for
a nonzero strangeness content of the nucleon~\cite{neu}.

In the future we will employ the present relativistic RPA formalism
to calculate the inclusive cross section for atmospheric neutrinos.
Atmospheric neutrinos have been observed over a wide range of energies
(from a few hundred MeV to several GeV) in large water \v{C}erenkov
detectors --- hence the need for a formalism that can address neutrino
physics over this broad energy range. Atmospheric neutrinos are particularly
interesting because of the current flavor anomaly in
the $\nu_\mu / \nu_e$ ratio.
This anomaly might signal neutrino oscillations. However, first one must
rule out all conventional nuclear effects. Thus, it
is important to examine these
nuclear-structure corrections before a definitive statement about new physics
can be made.

\acknowledgments

We would like to thank David K. Griegel
for careful reading of the manuscript and valuable comments.
We also thanks to Stefan Schramm for various suggestions.
This research was supported by U.S. Department of Energy under
Grant No.\ DE-FG02-87ER-40365, DE-FC05-85ER250000 and DE-FG05-92ER40750.
\eject
\appendix
\section*{1}

We adopt the form factor parameterization used in Ref.~\cite{don92b}. First,
the electromagnetic form factors are written in terms of simple dipole forms:
\begin{equation}
G=(1+4.97 \tau )^{-2},\ \tau=-q^2/4M^2,
\end{equation}
\begin{equation}
F_1^{(p)}=[1+\tau (1+\lambda _p)]G/(1+\tau),
\end{equation}
\begin{equation}
F_2^{(p)}=\lambda _p G/(1+\tau),
\end{equation}
\begin{equation}
F_1^{(n)}=\tau \lambda _n (1-\eta)G/(1+\tau),
\end{equation}
\begin{equation}
F_2^{(n)}=\lambda_n (1+\tau \eta)G/(1+\tau).
\end{equation}
Here the anomalous moments are,
\begin{equation}
\lambda _p = 1.793,\ \lambda _n = -1.913,
\end{equation}
and
\begin{equation}
\eta = (1+5.6\tau)^{-1}.
\end{equation}
This parameterization is good for the neutron form factors provided
$\tau \ll 1$. The isovector form factors are given by
\begin{equation}
F_1=F_1^{(p)}-F_1^{(n)} , \,\,
F_2=F_2^{(p)}-F_2^{(n)} \, .
\end{equation}
The axial form factor $G_A$ is
\begin{equation}
G_A={1.26 \over (1-q^2/M_A^2)^2}\,,
\label{gax}
\end{equation}
and the pseudoscalar form factor is given by
\begin{equation}
F_p={2M G_A \over m_\pi^2-q^2} \, .
\end{equation}

\appendix
\section*{2}

Analytic expressions for the imaginary parts of the polarizations introduced
in Sec.~\ref{subsec:impulse} are
\begin{eqnarray}
{\rm Im}(\Pi^{vv}_L) &=& {q^2\over 2 \pi {\bf q}^3} [E_3+q_0 E_2+
{q^2 E_1\over 4}]\ ,\\
{\rm Im}(\Pi^{vv}_T) &=& {q^2 \over 4 \pi {\bf q}^3} \biggr[ E_3
+q_0 E_2+\biggl({{\bf q}^2
M^{*2}\over q^2} + {q_0^2+{\bf q}^2 \over 4} \biggr)\ E_1 \biggl] \ ,\\
{\rm Im}(\Pi^{tt}_L) &=& {-q^2 \over 8 \pi {\bf q}^3 M^2}
\biggr[q^2 E_3+q_0 q^2
E_2 + (M^{*2} {\bf q}^2 + {q^2 q_0^2 \over 4})\ E_1 \biggl]\ ,\\
{\rm Im}(\Pi^{tt}_T) &=& {q^2 \over 16 \pi {\bf q}^3 M^2}
\biggr[(M^{*2}{\bf q}^2 -
{q^4 \over 4})E_1 - q_0 q^2 E_2 - q^2 E_3 \biggl]\ ,\\
{\rm Im}(\Pi_A) &=& {M^{*2} E_1 \over 2 \pi |{\bf q}|}\ ,\\
{\rm Im}(\Pi^{vt}_L) &=& {- q^2 M^* E_1 \over 8 \pi |{\bf q}| M } = -{\rm
Im}(\Pi^{vt}_T) \ ,\\
{\rm Im}(\Pi_{va}) &=& {q^2 \over 8 \pi {\bf q}^3}[2 E_2 + q_0 E_1]\ ,\\
{\rm Im}(\Pi_{pp}) &=& {q^2 E_1 \over 8 \pi |{\bf q}|}\ ,\\
{\rm Im}(\Pi_{ap}) &=& -{M^* E_1 \over 4 \pi |{\bf q}|}\ ,
\end{eqnarray}
where
\begin{eqnarray}
E_n= {E_F^n-E_-^n \over n}~~~~(n=1,2,3)
\end{eqnarray}
with
\begin{eqnarray}
E_F &=& \sqrt {k_F^2 + M^{*2}}\ ,\\
E_- &=& {\rm Min}(E_F, E_{max})\ ,\\
E_{max} &=& {\rm Max} \Biggl[ M^*, E_F-q_0, {1\over 2}\biggl(-q_0+|{\bf q}|
\sqrt {1-{4 M^{*2} \over q^2}}\ \biggr) \Biggr ]\ .
\end{eqnarray}
The vacuum part does not contribute to the impulse response for
spacelike momenta.

\begin{figure}
\caption{Feynman diagram for charged-current neutrino scattering. The
neutrino with momentum $k$ scatters off a nucleus in a state $\left| \psi_i
(p_i) \right\rangle$ and produces a muon or an electron with momentum $k'$.}
\label{scatt}
\end{figure}

\begin{figure}
\caption{A diagrammatic representation of the RPA correction
${\bf \Delta \Pi_{RPA}}$. The RPA propagator
${\bf D_{RPA}}$ is shown in (b)  }
\label{rpafig}
\end{figure}
\begin{figure}
\caption{Double differential cross sections for 1 GeV neutrinos.  The solid
curve is the momentum dependent calculation.  Figure (a) is for $|{\bf q}| =
$ 0.5 GeV while (b) is for $|{\bf q}| = $ 1.2 GeV.
The dashed (dot-dashed) curve is
the cross section obtained from a MFT (free Fermi gas) calculation.  }
\label{qu1}
\end{figure}
\begin{figure}
\caption{Double differential cross section with RPA correlations for
$|{\bf q}|=500$ MeV and $E_\nu=1 $ GeV.  The solid curve is the RPA
calculation with $M$ while the dashed curve is the RPA with $M^*$.
(a) is for $g'$=0, (b) is for $g'$=0.3, and (c) is for $g'$=0.7.
In (c) the $M$ and $M^*$ impulse calculations are also shown for
comparison.}
\label{qu2}
\end{figure}

\begin{figure}[h]
\caption{$d\sigma/dQ^2$ for a neutrino (a) and antineutrino (b)
scattering at an incoming energy of 1.2 GeV. The solid curve is
the impulse with $M$ calculation while the dashed curve is the
$M^{*}$ RPA with $g'=0.7$. The effect of the momentum-dependent
self-energies is indicated with the dot-dashed line.}
\label{Neu}
\end{figure}

\begin{figure}[h]
\caption{$d\sigma/dQ^2$ averaged over the BNL spectrum.  In (a) the
two solid curves denote the impulse and the RPA ($g'=0.7$) results
using the free nucleon mass $M$. Also shown (two dashed lines) are
the corresponding results using an effective nucleon mass $M^*$.
In (b) the cross sections are shown over the available experimental
range in $Q^2$. The solid and the dashed curves are the results of
impulse calculations with $M$ and $M^*$, respectively. The dot-dashed
is the calculation with momentum dependent self-energies. }
\label{bnlfig}
\end{figure}

\begin{table}
\caption{The ratio of cross section ${d \sigma \over d Q^2} (Q^2=0.2\,
{\rm GeV}^2)/ {d \sigma \over d Q^2}
(Q^2=1.0\, {\rm GeV}^2)$ for various nuclear structure effects at a given
axial mass $M_A$ in the first column. }

\begin{center}
\begin{tabular}{|c|c|c|c|c|c|}
\hline
&\multicolumn{5}{c|}{${d \sigma \over d Q^2} (Q^2=0.2\,{\rm GeV}^2)/
{d \sigma \over d Q^2}(Q^2=1.0\,{\rm GeV}^2)$}\\
\hline
 $M_A$ (GeV) & Impulse $M$ & Impulse $M^*$ &  RPA M$^*$ &
RPA M &$M_{\bf p}^{*}$ \\
\hline\hline
1.09 & 19.42 & 25.83 & 23.72 & 17.21& 21.85  \\
1.14& 18.47 &24.71& 22.59 &16.32& 20.83 \\
1.18& 17.75 & 23.78 & 21.69 & 15.66 &20.03  \\
1.21& 16.91 & 22.66& 20.61 &14.89& 19.08 \\
1.25& 15.95 & 21.36 & 19.38 & 14.02 & 18.00 \\
1.30& 14.9 & 19.92& 18.04 &13.07& 16.81 \\
\hline
\end{tabular}
\end{center}
\label{compa}

\end{table}
\begin{table}
\caption{The same as Table~\protect{\ref{compa}} for the ratios
${d \sigma \over d Q^2} (Q^2=0.5\, {\rm GeV}^2)/ {d \sigma \over d Q^2}
(Q^2=1.0\, {\rm GeV}^2)$. }
\begin{center}
\begin{tabular}{|c|c|c|c|c|c|}
\hline
&\multicolumn{5}{c|}{${d \sigma \over d Q^2} (Q^2=0.5\,{\rm GeV}^2)/
{d \sigma \over d Q^2}(Q^2=1.0\,{\rm GeV}^2)$}\\
\hline
 $M_A$ (GeV) & Impulse $M$ & Impulse $M^*$ &  RPA $M^*$ &
RPA M &$M_{\bf p}^{*}$\\
\hline\hline
1.09 & 5.47 & 6.50 & 6.40 & 5.34 & 5.77  \\
1.14 & 5.35 &  6.38 & 6.27 & 5.21 & 5.65 \\
1.18& 5.24 & 6.26 & 6.14 & 5.10 & 5.54  \\
1.21& 5.10 & 6.09& 5.96 &4.96& 5.39 \\
1.25& 4.92 & 5.88 & 5.75 & 4.78 & 5.21 \\
1.30& 4.72 & 5.62 & 5.49 &4.58& 4.99 \\
\hline
\end{tabular}
\end{center}
\label{compa1}

\end{table}

\end{document}